\documentclass[aps,prd,twocolumn,
preprintnumbers]{revtex4}
\usepackage{graphicx}
\def \s{\sqrt{2}}

\def\be{\begin{equation}}
\def\ee{\end{equation}}
\def\bea{\begin{eqnarray}}
\def\eea{\end{eqnarray}}
\def\bean{\begin{eqnarray*}}
\def\eean{\end{eqnarray*}}
\def\bary{\begin{array}}
\def\eary{\end{array}}

\def\bit{\begin{itemize}}
\def\eit{\end{itemize}}
\def\bwt{\begin{widetext}}
\def\ewt{\end{widetext}}
\def\half{\frac{1}{2}}

\def\ol{\overline}

\begin{document}

\title{Determination of $\gamma$ from charmless
$B^{\pm} \to M^0 M^{\pm}$ decays using U-spin}

\author{Amarjit Soni}
\author{Denis A.~Suprun}
\affiliation{High Energy Theory Group, 
Brookhaven National Laboratory, Upton, NY 11973}

\date{\today}

\begin{abstract}

U-spin multiplet approach is applied to the full set of charmless hadronic $B^{\pm} \to M^0 M^{\pm}$ decays for the purpose of precise extraction of the unitarity angle $\gamma$. Each of the four data sets, $P^0 P^{\pm}$, $P^0 V^{\pm}$, $V^0 P^{\pm}$ and $V^0 V^{\pm}$, with $P \equiv$ pseudoscalar and $V \equiv$ vector, can be used to yield a precise value of $\gamma$. The crucial advantage of this method over the common SU(3) symmetry based quark-diagrammatic approach is that no assumptions regarding relative sizes of topological decay amplitudes need to be made. As a result, this method avoids an uncontrollable theoretical uncertainty that is related to the neglect of some topological diagrams (e.g., exchange and annihilation graphs) in the SU(3) approach. Application of the U-spin approach to the current data yields: $\gamma=\left(54^{+12}_{-11}\right)^{\circ}$. We find that improved measurements of $\phi \pi^{\pm}$ and $\ol K^{*0} K^{\pm}$ branching ratios would lead to appreciably better extraction of $\gamma$. In this method, which is completely data driven, in a few years we should be able to obtain a model independent determination of $\gamma$ with an accuracy of O(few degrees). 
\end{abstract}


\maketitle

Precise determinations of the angles of the unitarity
triangle (UT) remains an important but difficult goal in Particle Physics.
Though methods for direct determinations of all the angles are
now known, we are still quite far away from having large enough sample
of B's that are needed~\cite{schune_ichep05}.
The main challenge in extracting the angles
from the data is of course that weak decays  take place in the
presence of strong interactions (i.e. QCD) which in this energy
regime has important, non-perturbative effects. Fortunately, QCD
respects flavor symmetries. Use of these symmetries presents
an important avenue to extract results, though often at the expense
of some accuracy. In the context of the angle
$\gamma$ of the UT, in fact SU(3) flavor symmetry has already been put to
very good use~\cite{Chiang:2004nm,Chiang:2003pm}. 
In this paper we propose to use U-spin
for determining $\gamma$ from charmless $B^{\pm}$ decays. 


There are substantial differences between U-spin multiplet approach and other phenomenological methods, such as SU(3) based approach, of understanding the current $B$ decay data.

\begin{itemize}
\item U-spin {\it multiplet} method has the significant advantage that, 
unlike SU(3) fits to charmless $B$ decays, quark diagrammatic {\it topological} approach is not invoked at all. Thus, we do not need to make any assumptions about the relative sizes of various contributing topological diagrams and so no amplitude need be neglected~\cite{Fleischer:1999pa}.

\item U-spin is a flavor symmetry similar to isospin. Since it is a subgroup of SU(3) flavor symmetry, we expect it to be at least as accurate as SU(3) and possibly better.

\item 
It is important to emphasize that noticeable flavor symmetry breaking effects in decay amplitudes do not necessarily lead to large uncertainties in $\gamma$ extraction. For instance, SU(3) breaking effects of about 20\% that are related to the ratio of decay constants $f_K$ and $f_{\pi}$, only lead to a small ($2^{\circ}$, or 3\%) theoretical uncertainty in determination of $\gamma$ from SU(3) fits~\cite{Suprun:2005}. The effects of the $\eta-\eta'$ mixing on the theoretical error in $\gamma$ were also found to be small ($\lesssim1^{\circ}$)~\cite{Chiang:2004nm}. Since in the U-spin approach graphical topologies are not used, estimate of U-spin breaking effects on $\gamma$ extraction may be amenable to calculational frameworks such as QCD factorization, pQCD, soft collinear effective theory (SCET), or QCD sum rules~\cite{effective-theories}.

\item The fact that the U-spin approach does not make use of graphical topologies, of course means that electroweak penguins are automatically fully contained within this approach. This is in sharp contrast to the case of $\alpha$ extraction with the use of isospin. Therein electroweak penguins complicate the extraction of the CKM phase $\alpha$ and present a serious limitation. In this respect the U-spin approach has an advantage over even the isospin method.

\item Needless to say, the standard $B \to D K$ methods of direct $\gamma$ extraction are theoretically the cleanest (error of O(.1\%)~\cite{schune_ichep05}) and should ultimately provide the most accurate determination of $\gamma$. But this accuracy will only be attained after very large data samples become available, perhaps many years down the road. The U-spin approach that we are using here, on the other hand, can provide a fairly accurate value of $\gamma$ (error of O(few percent)) with modest increase of luminosities.

\item Furthermore, since the U-spin approach automatically includes all penguin contributions, whereas the $B \to D K$ method do not, a comparison of the values of $\gamma$ from the two methods provides a crucial test for new physics.
\end{itemize}

We will show that there are four separate sets of two-body decays of charged B's each of which can give a value of $\gamma$. Existing data already allows determination of $\gamma$ with an accuracy
in the same ball park as other methods being used. We 
identify modes ($\phi \pi^{\pm}$ and 
$\ol K^{*0} K^{\pm}$) whose
improved experimental measurements should appreciably
improve the accuracy on $\gamma$ with this method. In the era of
the current B-factories, with the planned luminosities
of a few $ab^{-1}$, the method should allow us to determine
$\gamma$ with an accuracy of a few degrees. Furthermore,
as better experimental information, at these luminosities,
becomes available for
all the relevant data sets, this  method should give an
understanding of its inherent systematic error.

Let us very briefly recapitulate some elementary aspects of 
U-spin~\cite{Chiang:2003pm,Chiang:2003rb,recents,Fleischer:1999pa}.    
Recall that the U-spin subgroup of $SU(3)$ is 
similar to the I-spin (isospin) subgroup except that the quark doublets with $U = 1/2, U_3 = \pm 1/2$ are
\be
\label{eqn:qks}
\left[ \begin{array}{c} |\half~~\half \rangle \\
                               |\half -\! \half \rangle \end{array}
\right] = \left[ \begin{array}{c} |d \rangle \\ |s \rangle \end{array}
\right], \quad
\left[ \begin{array}{c} |\half~~\half \rangle \\
                                      |\half -\!\half \rangle \end{array}
\right] = \left[ \begin{array}{c} |\bar s \rangle \\ -\!| \bar d \rangle
\end{array} \right].
\ee

$B^+$ is a U-spin singlet, while charged charmless mesons $\pi^+ (\rho^+)$, $K^+ (K^{*+})$ belong to U-spin doublets. Neutral mesons may get contributions from a U-spin triplet and two U-spin singlets. Strange neutral mesons  $K^0 (K^{*0})$ and their antiparticles are pure U-spin triplets.
The multiplet decompositions of other neutral mesons can be determined to be
\bea
\label{eqn:P-neutrals}
\begin{array}{lll}
\pi^0 & = & -\frac12 |1~~0 \rangle + \frac{\sqrt3}{2} |0~~0 \rangle_8~~,\\
\eta & = & \sqrt{\frac23} |1~~0 \rangle + \frac{\sqrt2}{3} |0~~0 \rangle_8 - \frac13 |0~~0 \rangle_1~~,\\
\eta' & = & \frac{1}{2\sqrt3} |1~~0 \rangle + \frac16 |0~~0 \rangle_8 + 
\frac{2\sqrt 2}{3} |0~~0 \rangle_1~~, \\
\rho^0 & = & -\frac12 |1~~0 \rangle + \frac{\sqrt3}{2} |0~~0 \rangle_8~~,\\
\omega  & = &  \quad -\frac12 |1~~0 \rangle - \frac{\sqrt3}{6} |0~~0 \rangle_8 + \sqrt\frac23 |0~~0 \rangle_1~~,\\
\phi  & = &  \frac{1}{\sqrt 2} |1~~0 \rangle + 
\frac{1}{\sqrt 6} |0~~0 \rangle_8 + \frac{1}{\sqrt 3} |0~~0 \rangle_1~~,
\end{array}
\eea
where two U-spin singlets are defined as
\bea
\begin{array}{lll}
|0~~0 \rangle_8 & \equiv & \frac{1}{\sqrt{6}} | s \bar s + d \bar d - 2u \bar u \rangle ~~,\\
|0~~0 \rangle_1 & \equiv & \frac{1}{\sqrt{3}} | u \bar u + d \bar d + s \bar s \rangle  ~~.
\end{array}
\eea 

One may decompose the $|\Delta S|=1$ and $\Delta S=0$ effective Hamiltonians
into members of {\em the same} two U-spin doublets multiplying given 
CKM factors.
For practical purposes, 
using CKM unitarity,
it is convenient to write the effective Hamiltonian
so that it involves only the u and c quarks:

\bea
\label{Hd}
\Delta S=0: \quad {\cal H}_{\rm eff}^{\bar b\to\bar d} & = & V^*_{ub}V_{ud}O^u_d + 
V^*_{cb}V_{cd}O^c_d~~,\\
\label{Hs}
|\Delta S|=1: \quad {\cal H}_{\rm eff}^{\bar b\to\bar s} & = & V^*_{ub}V_{us}O^u_s + 
V^*_{cb}V_{cs}O^c_s~~.
\eea
The assumption of U-spin symmetry implies that U-spin doublet 
operators $O^u_d$ and $O^u_s$ are identical, as well as 
the $O^c_d$ and $O^c_s$ operators. The subscripts $d$ and $s$ 
may be omitted. Hadronic matrix elements of these two 
operators, $O^u$ and $O^c$, will be denoted $A^u$ and $A^c$ and will 
be referred to as ``u-like" and ``c-like" amplitudes~\cite{tree_foot}, 
where the latter includes electroweak penguin contributions.  Note that these amplitudes multiply different CKM factors in $|\Delta S|=1$ and $\Delta S=0$ processes.

In isospin analysis of $B$ decays~\cite{Gronau:1990ka} the effective Hamiltonian transforms as either $\Delta I=\frac12$ or $\Delta I=\frac32$. 
While electroweak penguins,
violate isospin due to the charge difference between $u$ and $d$ quarks, they 
do not violate U-spin. There are only three topological diagrams that 
may contribute to charged $B$ decays: tree, penguin (QCD and electroweak), 
and annihilation. The effective Hamiltonian of any of these 
decay types transforms as a U-spin doublet, $\Delta U = \half$. 
This makes U-spin a particularly convenient approach that allows the 
complete description of charged $B$ decays without making additional 
assumptions on the size of individual topological diagrams and without 
neglecting any of them, including annihilation. While the SU(3) 
based approach~\cite{Chiang:2004nm,Chiang:2003pm} does not inherently 
require ignoring annihilation, exchange and penguin annihilation 
contributions, in practice one has to do that to limit the number 
of parameters and keep SU(3) fits stable. This advantage of the 
U-spin approach makes it particularly appealing; in the long run,
it should significantly reduce 
theoretical uncertainties associated with this method.

Since the initial $B^+$ meson is a U-spin singlet and the effective Hamiltonian always transforms like a U-spin doublet, the final $M^0 M^+$ states must be U-spin doublets. They can be formed in three different ways.
While the charmless charged meson $M^+$ can only belong to a doublet, 
the neutral meson $M^0$ can be a linear combination of three 
different multiplets. As a result, any $\Delta S = 0$, $B^+ \to M^0 M^+$ 
decay amplitude can be expressed in terms of three 
amplitudes: $A^d_1$, $A^d_0$, $B^d_0$. They correspond to the 
U-spin triplet, U-spin singlet $|0~~0\rangle_8$, and SU(3) 
singlet $|0~~0\rangle_1$ contributions into the decay amplitude. 
Each of these three amplitudes consists of a ``u-like" and a ``c-like" 
part (Eq.~\ref{eq:amps}). Similarly, any $|\Delta S| = 1$ decay 
amplitude can be written in terms of three other 
amplitudes: $A^s_1$, $A^s_0$, $B^s_0$. The assumption of U-spin 
symmetry implies that the difference between $A^d_1$ and $A^s_1$ 
comes only through the difference in the CKM matrix elements. Thus, 
the complete amplitudes for U-spin final states are 
given by

\bea
\label{eq:amps}
\begin{array}{llll}
\Delta S=0: & A^d_{0,1} & = & V^*_{ub}V_{ud}A^u_{0,1} + V^*_{cb}V_{cd}A^c_{0,1}~~,\\
 & B^d_0 & = &  V^*_{ub}V_{ud}B^u_0 + V^*_{cb}V_{cd}B^c_0~~,\\
|\Delta S|=1: &  A^s_{0,1} & = & V^*_{ub}V_{us}A^u_{0,1} + V^*_{cb}V_{cs}A^c_{0,1}~~,\\
& B^s_0 & = &  V^*_{ub}V_{us}B^u_0 + V^*_{cb}V_{cs}B^c_0~~.
\end{array}
\eea

Then we find that physical decay amplitudes for $V^0 P^+$ and $V^0 V^+$ 
modes may be decomposed into U-spin amplitudes~\cite{conv1_foot},

\bea
\begin{array}{llcc}
\label{eqn:KK}
A(\ol K^{*0} K^+), & A(\ol K^{*0} K^{*+}) & = &  - 2\s A^d_1~~,\\
A(\rho^0\pi^+), & A(\rho^0\rho^+) & = &  3 A^d_0 - A^d_1~~,\\
A(\omega\pi^+), & A(\omega\rho^+) & = &  -A^d_0 - A^d_1 + \s B^d_0~~,\\
A(\phi \pi^+), & A(\phi \rho^+) & = &  \s A^d_0 + \s A^d_1 + B^d_0~~,\\
A(K^{*0}\pi^+), & A(K^{*0}\rho^+) & = &  -2\s A^s_1~~,\\
A(\rho^0 K^+), & A(\rho^0 K^{*+}) & = &  3 A^s_0 + A^s_1~~,\\
A(\omega K^+), & A(\omega K^{*+}) & = &  -A^s_0 + A^s_1 + \s B^s_0~~,\\
A(\phi K^+), & A(\phi K^{*+}) & = &  \s A^s_0 -\s A^s_1 + B^s_0~~,
\end{array}
\eea
where $A_1,~A_0$ and $B_0$ correspond to final states with vector mesons $V^0$ in the U-spin triplet, in the octet U-spin singlet and in the SU(3) singlet, respectively. Naturally, the formulae for related $V^0 P^+$ and 
$V^0 V^+$ decay modes are the same, as seen in the above relations. However, the actual values for each of the U-spin amplitudes are constant only within each of the two subsets. They accept different values in $V^0 P^+$ and $V^0 V^+$ subsets.

Thus, eight $V^0 P^+$ decays are described by 12 parameters: six U-spin amplitudes $|A^u_{0,1}|$, $|A^c_{0,1}|$, $|B^u_0|$ and $|B^c_0|$, five relative strong phases between them and the weak phase $\gamma$. The same statement is separately valid for eight $V^0 V^+$ modes, too.

In the same way one can decompose physical amplitudes for $P^0 P^+$ and 
$P^0 V^+$ decay modes into U-spin amplitudes. 
Then we find~\cite{conv2_foot}:  
\bea
\begin{array}{llcc}
\label{eqn:Uspin}
A(\ol K^0 K^+), & A(\ol K^0 K^{*+}) & = &  - 2\s A^d_1~~,\\
A(\pi^0\pi^+), & A(\pi^0\rho^+) & = &  3 A^d_0 - A^d_1~~,\\
A(\eta\pi^+), & A(\eta\rho^+) & = &  \frac{2\s}{\sqrt3}A^d_0+\frac{2\s}{\sqrt3}A^d_1 - B^d_0~~,\\
A(\eta'\pi^+), & A(\eta'\rho^+) & = &  \frac{1}{\sqrt3} A^d_0+
\frac{1}{\sqrt3}A^d_1+2\s B^d_0~~,\\
A(K^0\pi^+), & A(K^0\rho^+) & = &  - 2\s A^s_1~~,\\
A(\pi^0 K^+), & A(\pi^0 K^{*+}) & = &  3 A^s_0 +  A^s_1~~,\\
A(\eta K^+), & A(\eta K^{*+}) & = &  \frac{2\s}{\sqrt3}A^s_0-\frac{2\s}{\sqrt3}A^s_1 - B^s_0~~,\\
A(\eta' K^+), & A(\eta' K^{*+}) & = &  \frac{1}{\sqrt3} A^s_0-
\frac{1}{\sqrt3} A^s_1+2\s B^s_0~~.
\end{array}
\eea
Just as the two subsets of $M^0 M^+$ that were considered before, $P^0 P^+$ and $P^0 V^+$ are also separately described by similar 12 parameters.

Charmless hadronic decays of the $B^+$ meson to the two-meson final states 
that contain vector $V$ or pseudoscalar $P$ mesons comprise 
four subsets: $P^0 P^+$, $V^0 V^+$, $V^0 P^+$, and $P^0 V^+$. 
Each of the subsets comprises eight decays, with all possible combinations of two charged mesons (e.g., $\pi^+$ and $K^+$ in the pseudoscalar octet) and four neutral ones ($K^{*0}$, $\rho$, $\omega$, and $\phi$ in the vector octet).
Thus there are altogether 16 relevant decays of $B^{\pm}$
of each of the four types.
Each of the subsets, again, is described by 12 parameters, namely, 6 U-spin amplitudes, 5 relative strong phases between them, and the weak phase $\gamma$ which is the only common parameter among four parameter sets. 

All 8 $B^+ \to P^0 P^+$ decays have actually been observed and their 
branching ratios and $CP$ asymmetries have been measured,
though, with the present statistics in most cases the errors
are rather large. This is especially so for the CP-asymmetries. 
In any case, with 16 data points and 12 fit parameters one can perform 
a fit and extract the preferred values for all parameters. 

In the other 3 subsets some modes have not yet been observed but upper limits on their branching ratios were reported. 
Needless to say, direct $CP$ asymmetries for these modes have not been 
determined yet. For some of these modes a central value and a large uncertainty 
are known. For the others, where only an upper limit at $90\%$ 
confidence level is reported, one can take central value as 
equal to 0 and approximately estimate the uncertainty by dividing the 
upper limit value by 2. 
For example, from ${\cal B}(B^+ \to \omega \rho^+)<16$ we 
crudely estimate that ${\cal B}(\omega \rho^+)=0.0\pm8.0$~\cite{units}. The data 
from upper limits helps in two ways. First of all, it provides 
additional data points, making a U-spin fit feasible. 
Second, it allows us to verify that the resulting fit is consistent with 
the current upper limits. 

In the case of $V^0 P^+$ decays, for instance, 6 out of 8 modes have been observed and provide 12 data points. The remaining two decays, $\ol K^{*0} K^+$ and $\phi \pi^+$, have not yet been observed. At present only the upper limits for these two modes are known: 
${\cal B}(B^+ \to \ol K^{*0} K^+)=0.0^{+1.3+0.6}_{-0.0-0.0} \; (<5.3)$~\cite{Jessop:2000bv} and 
${\cal B}(B^+ \to \phi \pi^+)<0.41$~\cite{Aubert:2003hz}. 
From these measurements we can crudely estimate that 
${\cal B}(B^+ \to \ol K^{*0} K^+)=0.00^{+1.43}_{-0.00}$ and
${\cal B}(B^+ \to \phi \pi^+)=0.0\pm0.2$. 
To make sure that the fit is consistent with the upper limits on the 
$\ol K^{*0} K^+$ and $\phi \pi^+$ branching ratios we add these two data points to the fit. Thus, the 12-parameter $V^0 P^+$ U-spin fit features 14 data points, making  $\gamma$ extraction possible.

Similarly, in the $V^0 V^+$ sector 5 modes have been detected 
and the first measurement of their $CP$ asymmetries has been 
attempted
(though, again, with rather large errors) for a total of 10 data points. 
The other 3 modes have not yet been observed but the upper 
limits were reported, allowing estimates of their branching ratios. 
The total number of $V^0 V^+$ data points rises to 13.

The least is known about $P^0 V^+$ decays. Not even an upper limit is known for $\bar{K^0} K^{*+}$. Of the remaining 7 decays modes only 4 have been detected, providing 8 data points. For the other three an estimate of the branching ratio can be made using current upper limits. Thus, there are only 11 data points and a reasonable 12 parameter U-spin fit cannot be performed. To avoid this problem, one can make a joint U-spin fit to two $M^0 M^+$ decay subsets, e.g.\ a fit to both $V^0 P^+$ and $P^0 V^+$ decays. With $\gamma$ being the only common parameter for both parameter sets, there are 11 completely free $P^0 V^+$ U-spin parameters (amplitudes and strong phases) that describe 11 $P^0 V^+$ data points. There is just enough data to make the joint fit work.

\begin{table*}[t]
\begin{center}
\small
\caption{Results of the U-spin fits to various subsets of charmless $B^{\pm} \to M^0 M^{\pm}$ decays. The bottom panel shows $\gamma$ as determined from direct measurements in $B \to D^{(*)} K^{(*)}$ decays, from indirect constraints on the apex of the unitarity triangle, and from SU(3) fits to charmless $PP$ decays.
\label{tab:fits}}
\begin{tabular}{clccc}
\hline
\hline
\quad Fit~\quad\quad & Subset & Modes &$\chi^2/dof$ & $\gamma$\\
\hline
1.& $V^0 P^+$ & 
$\ol K^{*0} K^+$\, $\rho^0\pi^+$\,$\omega\pi^+$\,$\phi\pi^+$\quad 
$ K^{*0} \pi^+$\,$\rho^0K^+$\,$\omega K^+$\,$\phi K^+$  & 
3.97/2 & $\left(30^{+17}_{-18}\right)^{\circ}$  \\
2.& $P^0 P^+$ & 
$\ol K^0 K^+$\, $\pi^0\pi^+$\,$\eta\pi^+$\,$\eta'\pi^+$\quad 
$ K^0 \pi^+$\,$\pi^0K^+$\,$\eta K^+$\,$\eta' K^+$ & 
3.01/4 & $\left(68^{+59}_{-14}\right)^{\circ}$  \\
3.& $V^0 V^+$ & 
$\ol K^{*0} K^{*+}$\, $\rho^0\rho^+$\,$\omega\rho^+$\,$\phi\rho^+$\quad 
$ K^{*0} \rho^+$\,$\rho^0K^{*+}$\,$\omega K^{*+}$\,$\phi K^{*+}$ & 
0.05/1 & $\left(40^{+136}_{-35}\right)^{\circ}$ \\
4A.& $P^0 V^+$ & 
$\ol K^0 K^{*+}$\, $\pi^0\rho^+$\,$\eta\rho^+$\,$\eta'\rho^+$\quad 
$ K^0 \rho^+$\,$\pi^0K^{*+}$\,$\eta K^{*+}$\,$\eta' K^{*+}$ 
& \multicolumn{2}{l}{\quad ~ insufficient data} \\
4B.& $(V^0 P^+ \bigcup P^0 V^+)$ & & 4.03/2 & $\left(30^{+17}_{-18}\right)^{\circ}$ \\
\hline
5.& \multicolumn{2}{l}{$(V^0 P^+ \bigcup P^0 P^+)$}  & 10.02/7 & $\left(54^{+12}_{-11}\right)^{\circ}$\\
6.& \multicolumn{2}{l}{$(V^0 P^+ \bigcup P^0 P^+ \bigcup V^0 V^+)$} & 10.47/9 & $\left(54^{+12}_{-11}\right)^{\circ}$  \\
\hline
\hline
\multicolumn{3}{l}{Direct measurements, BaBar~\cite{Aubert:2005yj}}
& & $(67\pm28\pm13\pm11)^{\circ}$ \\
\multicolumn{3}{l}{Direct measurements, Belle~\cite{Abe:2004gu}} 
& & $\left(68^{+14}_{-15}\pm13\pm11\right)^{\circ}$ \\
\multicolumn{3}{l}{Indirect constraints, CKMFitter~\cite{Charles:2004jd}} & & $\left(57.3^{+7.3}_{-12.9}\right)^{\circ}$\\
\multicolumn{3}{l}{Indirect constraints, UTFit~\cite{Bona:2005vz}} & & $\left(57.9\pm7.4\right)^{\circ}$\\
\multicolumn{3}{l}{SU(3) fits to $VP$ decays~\cite{Suprun:2005}} & & $\left(66.2^{+3.8}_{-3.9}\pm0.1\right)^{\circ}$\\
\multicolumn{3}{l}{SU(3) fits to $PP$ decays~\cite{Suprun:2005}} & & $\left(59\pm9\pm2\right)^{\circ}$\\
\hline
\hline
\end{tabular}
\end{center}
\end{table*}

Table~\ref{tab:fits} shows the results of the U-spin fits to four subsets of $M^0 M^+$ decays and their combinations. The top part of the table shows three fits to individual subsets ($V^0 P^+$, $P^0 P^+$, $V^0 V^+$) and one joint fit. As was mentioned before, the only way to explore $P^0 V^+$ data  is to make a joint fit, for example, the $(V^0 P^+ \bigcup P^0 V^+)$ one.  

Among the first four fits, the $V^0 P^+$ one stands out. It is the 
only fit that features a deep minimum at its preferred value of $\gamma$, 
with both upper and lower uncertainties staying under $20^{\circ}$. 
The $P^0 P^+$ fit has, on the other hand, a shallow minimum with 
very large upper uncertainty. Of the other two U-spin fits in 
the top part of the table, the $V^0 V^+$ one
produces a very shallow minimum, leaving $\gamma$ practically undetermined. 
As was mentioned before, the current data on $P^0 V^+$ is 
insufficient for a U-spin fit. Instead, $P^0 V^+$ subset was 
combined with the other $VP$ decays ($V^0 P^+$), making the joint 
fit possible. However, the effect of the $P^0 V^+$ data appears 
to be insignificant; the joint fit produces practically identical 
results to those of the $V^0 P^+$ fit. One can draw the 
conclusion that the quality of the current
data for $P^0 V^+$ and $V^0 V^+$ is unlikely 
to significantly affect joint fits. 

This is confirmed in the lower part of the table. The best U-spin fit is achieved when $V^0 P^+$ and $P^0 P^+$ data are combined (30 data points) and fitted with 23 parameters (two sets of six U-spin amplitudes and five strong phases, plus the weak phase $\gamma$). The joint $(V^0 P^+ \bigcup P^0 P^+)$ fit prefers a value of $\gamma$ that is in between the values favored by $V^0 P^+$ and $P^0 P^+$ fits, namely, $\gamma=(54^{+12}_{-11})^{\circ}$. The addition of the $V^0 V^+$ subset does not change this result, as expected. 

The above results are based on the latest world average values for branching ratios and $CP$ asymmetries in charged charmless $B$ decays~\cite{HFAG}. When the individual values from BaBar and Belle are very different, we employed the PDG scaling factor $S$ to boost uncertainties on the weighted averages. This modification only slightly affects the final result. The joint U-spin $(V^0 P^+ \bigcup P^0 P^+)$ fit to the {\it unscaled} data prefers the same central value but slightly smaller uncertainties for the weak phase: $\gamma=(54^{+11}_{-10})^{\circ}$.

We also explored the joint $(V^0 P^+ \bigcup P^0 P^+)$ fit in some detail with the purpose of estimating the expected improvement of $\gamma$ extraction as higher statistics on $B$ decays get accumulated.
We tried to identify some specific modes where smaller uncertainties 
on branching ratios would help reduce the error on $\gamma$. 

We found that setting a stricter upper limit on 
the $\phi \pi^+$ branching ratio is of particular importance. 
The current upper 
limit ${\cal B}(B^+ \to \phi \pi^+)<0.41$~\cite{Aubert:2003hz} is 
based on $89$ million $B\bar{B}$ pairs. Both $B$ factories will 
each accumulate in excess of $500$ million $B\bar{B}$ pairs 
by the summer of 2006. The available statistics on 
${\cal B}(\phi \pi^+)$ decays will increase by about a factor of 10, 
leading to uncertainties that are about 3 times smaller than 
the current ones. With the new data point 
of ${\cal B}(\phi \pi^+)=0.00\pm0.07$ the joint 
$(V^0 P^+ \bigcup P^0 P^+)$ U-spin fit features a 
rather deep minimum with uncertainties on $\gamma$ at the level 
of $8^{\circ}$. The improvements in $\gamma$ extraction due to stricter upper limits on $\ol K^{*0} K^+$  are somewhat smaller.

Finally, we scaled down uncertainties in all data points to the levels 
corresponding to 1 billion $B\bar{B}$ pairs. The U-spin fit becomes 
much deeper at its minimum and $\gamma$ gets extracted with 
a $6^{\circ}$ uncertainty. Theoretical uncertainties associated with 
this method are expected to be small so it has the potential to put 
rather stringent constraints on the weak phase $\gamma$~\cite{neutrals_foot}.

Summarizing, with current statistics, the best U-spin fits allow the 
extraction of $\gamma$ with a reasonable accuracy and the 
preferred value is $\gamma=(54^{+12}_{-11})^{\circ}$ which is quite  
consistent with the current 
indirect determinations that expect $\gamma$ to lie 
between $42^{\circ}$ and $73^{\circ}$ \cite{Charles:2004jd,Bona:2005vz}. 
Note that the intrinsic theoretical uncertainty associated with 
possible U-spin breaking effects is expected to be rather small and that U-spin symmetry is the only assumption that is made 
in this approach~\cite{Gronau:2004tm}. 
Clearly, as data with higher statistics becomes available, the 
statistical uncertainties on $\gamma$ will become even smaller. 
At the moment the difference between the four values of $\gamma$ 
extracted from the four subsets is not very meaningful due to 
large uncertainties (Table~\ref{tab:fits}). When all branching 
ratios and $CP$ asymmetries in charged $B$ decays are 
experimentally determined with high accuracy, U-spin approach 
should enable extraction of  $\gamma$ quite precisely from 
each of the four subsets of data. 
The resulting spread in $\gamma$ values should be small and could perhaps be used to indicate the systematic errors inherent in the method due to residual U-spin breaking effects. The crucial advantage of the method is that the extraction of $\gamma$ is completely model independent and entirely data driven. Note also that unlike the use of isospin for $\alpha$, electroweak penguins are not a problem in our approach.
Penguin contributions are entering in an important way in this U-spin approach for getting $\gamma$. That means that this method is
sensitive to new physics in the loops. In contrast, recall
that the standard $B \to DK$ methods~\cite{schune_ichep05} involve only tree $B$ decays. Comparison of $\gamma$ from these two methods is therefore important for uncovering new physics.  

We thank J.~G.~Smith for helpful discussions. This research was supported in part by DOE contract Nos.DE-FG02-04ER41291(BNL).


\begin{thebibliography}{99}
\bibitem{tree_foot}
Since both the u and the c piece in the $H_{eff}$ also
involve pieces of the t-quark, use of tree and penguin 
terminology for these, which is sometimes done, may be misleading.
\bibitem{units}
All branching ratios in this paper are in units of $10^{-6}$.
\bibitem{neutrals_foot}  
 The work on neutral $B$'s is in progress.
\end{thebibliography}
\end{document}